%% file: template.tex
\documentclass[compactaffiliation]{Interspeech}
\usepackage{multirow}
\usepackage{tipa}
\interspeechcameraready

\title{Intelligibility of Text-to-Speech Systems for Mathematical Expressions \\
\small{Accepted at Interspeech 2025.}
}

\author[affiliation={1}]{Sujoy}{Roychowdhury}
\author[affiliation={1}]{Ranjani}{H.G.}
\author[affiliation={1}]{Sumit}{Soman}
\author[affiliation={2}]{Nishtha}{Paul}
\author[affiliation={1}]{Subhadip}{Bandyopadhyay}
\author[affiliation={3}]{Siddhanth}{Iyengar}

\affiliation{}{Ericsson R\&D, Bangalore}{India} 
\affiliation{}{IIIT Bangalore}{India}
\affiliation{}{BITS Pilani}{India}
\email{\{sujoy.roychowdhury, ranjani.h.g, sumit.soman, subhadip.bandyopadhyay\}@ericsson.com}
\keywords{TTS, mathematical expressions, \LaTeX, listening tests, perception experiments, large language models.}

\usepackage{comment}
\usepackage{amsmath}
\usepackage{algorithm}
\usepackage{algpseudocode}
\usepackage[symbol]{footmisc}

\begin{document}

\maketitle
% \footnote[7]{Accepted at Interspeech 2025.}
% the abstract here must exactly match the abstract entered into the paper submission system
\begin{abstract}
There has been limited evaluation of advanced Text-to-Speech (TTS) models with Mathematical eXpressions (MX) as inputs. In this work, we design experiments to evaluate quality and intelligibility of five TTS models through listening and transcribing tests for various categories of MX. We use two Large Language Models (LLMs) to generate English pronunciation from LaTeX MX as TTS models cannot process LaTeX directly. We use Mean Opinion Score from user ratings and quantify intelligibility through transcription correctness using three metrics. We also compare listener preference of TTS outputs with respect to human expert rendition of same MX. Results establish that output of TTS models for MX is not necessarily intelligible, the gap in intelligibility varies across TTS models and MX category.  For most categories, performance of TTS models is significantly worse than that of expert rendition. The effect of choice of LLM is limited. This establishes the need to improve TTS models for MX.
\end{abstract}

%{\color{blue}- \LaTeX\ Character Error Rate (LCER), count-of-correctness via manual verifications and TeXBLEU}

\section{Introduction}\label{sec:intro} 
There have been significant advances in Text-to-Speech (TTS) models alongside those of Large language models (LLMs) in the recent years \cite{cui2024recent, barakat2024deep, e2tts2024}. These have resulted in high quality TTS outputs that are nearly indistinguishable from human speech \cite{barakat2024deep, e2tts2024}. Typical applications for expressive speech consider and evaluate audio versions of eBooks or podcasts. However, there has been limited (and recent) attention to audio versions of scientific text based content. It typically includes Mathematical eXpressions (MX), tabular data, graphs, or block diagrams, which have higher complexity than conversational or textual content alone. {\color{black} In particular, for} TTS MX outputs, it has been established that existing objective TTS evaluation metrics, such as Word Error Rate (WER) with respect to the original transcriptions, does not suffice \cite{mathbridge}. 
% and speaker similarity \cite{kamil2022automatic} is not relevant as evaluation is not on speaker characteristics here. 

{\color{black} It is expected that TTS output (referred to AudioMX, henceforth) for MX input is such that a listener can discern and hence transcribe the MX. To the best of our knowledge, there has been no work which addresses perception and intelligibility aspects of TTS outputs for MX through listening experiments.} We design experiments to systematically understand challenges {\color{black} of TTS models in converting MX to AudioMX and their interpretation by human listeners}. We present both subjective and objective metrics to evaluate human interpretability captured through transcription in \LaTeX. %including Mean Opinion Score (MOS) scores for subjective evaluation through listening tests; we measure intelligibility through human interpretation and transcription of the AudioMX  - \LaTeX\ Character Error Rate (LCER) of the \LaTeX\ transcription and TeXBLEU \cite{jung2024texbleu}, 
We use outputs of five commercially available and popular TTS models (which are considered as closed box or opaque systems, in this work). 
%have resulted in advances to mitigate some of the latency challenges in cascading LLMs with traditional ASR or text-to-speech systems. 

\subsection{Existing work}\label{subsec:LitReview} 

One of the earliest known MX related datasets is Handwritten and Audio Dataset of Mathematical Expressions (HAMEX) \cite{handwritten_math_expressions}, which associated audio and handwritten MX. It was used to consider audio as an additional modality for recognition of handwritten equations \cite{medjkoune2017combining} and not for the TTS problem. TTS for MX is challenging and earlier works \cite{caky2009mathematical,bier2015adaptive, bier2019rule, mogale2020grammar} utilized rules and heuristics for converting MX into text of the spoken language (Polish in \cite{bier2015adaptive} and Sepedi in \cite{mogale2020grammar}), followed by a TTS model for the speech outputs - the TTS has not been evaluated. 

\textit{MathSpeech} \cite{hyeon2024mathspeech} developed a benchmark dataset and an approach to transcribe MX from audio (extracted from YouTube video lectures) to \LaTeX~ using fine-tuned small Language Models; this involved correcting Automatic Speech Recognition (ASR) errors. \textit{MathBridge} \cite{mathbridge} introduced a dataset of spoken MX and their corresponding \LaTeX~ representations, benchmarked on T5-large \cite{2020t5} using sacreBLEU \cite{post-2018-call} as a metric. Speech to \LaTeX\ challenges are discussed along two approaches -  LLMs as post-processing step to ASR, and multimodal LMs for English and Russian languages in \cite{korzh2024listening}. All of these are ASR systems for transcribing MX from speech.

TTS for MX has been proposed and evaluated in \textit{Mathreader }\cite{hyeon2025mathreadertexttospeechmathematical} and \textit{MathVision} \cite{awais2024mathvision}. The former uses the \textit{MathBridge} dataset to fine-tune T5 model to convert \LaTeX~ to Spoken English sentences. However, it assumes TTS models considered are error-free \cite{hyeon2025mathreadertexttospeechmathematical}. %This is 
In \textit{MathVision}, YOLO v7 vision models classify MX to categories, followed by LSTM models to convert output of YOLO models to a sentence in natural language. However, we note that both these works assume TTS systems are error free, and do not evaluate the TTS systems. {\color{black} The need for prosody in TTS outputs for MX inputs has been observed in \cite{souza2020towards}}.  

%Evaluating TTS systems involving \LaTeX~ expressions also requires metrics such as TeXBLEU \cite{jung2024texbleu}. %Some of the challenges in using TeXBLEU are highlighted in Section \ref{subsec:texBleu}.
% our work ({\color{red}{refer Table \ref{tab:texbleu}}}).

In this work, we focus on intelligibility and perceived quality of some of the popular TTS models through listening tests. A major distinction from the existing works and ours is that we focus on evaluating commercial, state-of-the-art (SOTA) TTS models \cite{LabelboxTTS2024} via MOS and LCER. We study the effect of choice of TTS models for various categories of MX. {\color{black}We now present our Research Questions (RQs) and key contributions.} %Section \ref{subsec:RQ} lists the research questions}. 

 %{\color{blue}{Place our work in this context}}

\subsection{Research Questions and Contributions}\label{subsec:RQ}

The objective of TTS models for MX is that the output audio conveys the content in an intelligible way to the listener.  It is critical to be able to follow and comprehend MX correctly. {\color{black}Note that this is not same as cascaded  system of TTS with ASR} (discussed in Section \ref{subsec:asrCheck}). We consider the following RQs.
\begin{itemize}
    % \item \textbf{RQ1}:- Can a listener replicate the MX by listening to the TTS output audio alone and without access to the MX
    % \item \textbf{RQ1a}:- {\color{black}{Does the comprehension of MX from a TTS audio depend on the category of MX}}
    % \item \textbf{RQ1}:- Does a listener's ability to transcribe an MX correctly from the audio rendition depend on the MX category
    \item \textbf{RQ1}:- Can we measure a listener's perception of a TTS audio output of a MX? Does their ability to transcribe it correctly depend on the MX category?
    \item \textbf{RQ2}:- Is there any perceived difference between an human expert rendered audio recording of a MX (RAudioMX) and TTS output MX (AudioMX) in terms of intelligibility, quality, clarity and/or speed?
\end{itemize}

To the best of our knowledge, this is the first study of the TTS models for MX with a focus on user perception and intelligibility. Our \textbf{main contributions} are:
\begin{itemize}
    \item Establish that the intelligibility of TTS outputs varies with the MX category, and that the relative performance of TTS models is dependent on the type of the MX.
    \item Demonstrate that there is a gap between user's perception of their understanding of the MX and the ability to transcribe it.
    % \item Establish that except for numerical expressions and fractions, there is gap between TTS models and reference rendering by experts of MX.
    \item Identify categories of MX where there is a gap between TTS outputs and reference rendering by experts.
    \item Qualitatively show that challenges in intelligibility are related to prosody and duration of the synthesis of (parts of) MX. 
\end{itemize} 

% {\color{blue}We design L1 and L2 listening tests, described in Section \ref{sec:exp_design}} to address our RQs.

%{\color{blue}The paper is structured as follows: Section \ref{sec:methodologyAndDataset} elaborates on the dataset considered along with proposed methodology (illustrated in Figure \ref{fig:exptdesign}), Section \ref{sec:exp_design} details design of two listening tests,  followed by Section \ref{sec:results} where we analyze the results of the experiments. A qualitative analysis of the output of the experiments is discussed in Section \ref{sec:discussion}} followed by concluding section. 

\begin{figure}
    \centering
    \includegraphics[width=0.92\linewidth]{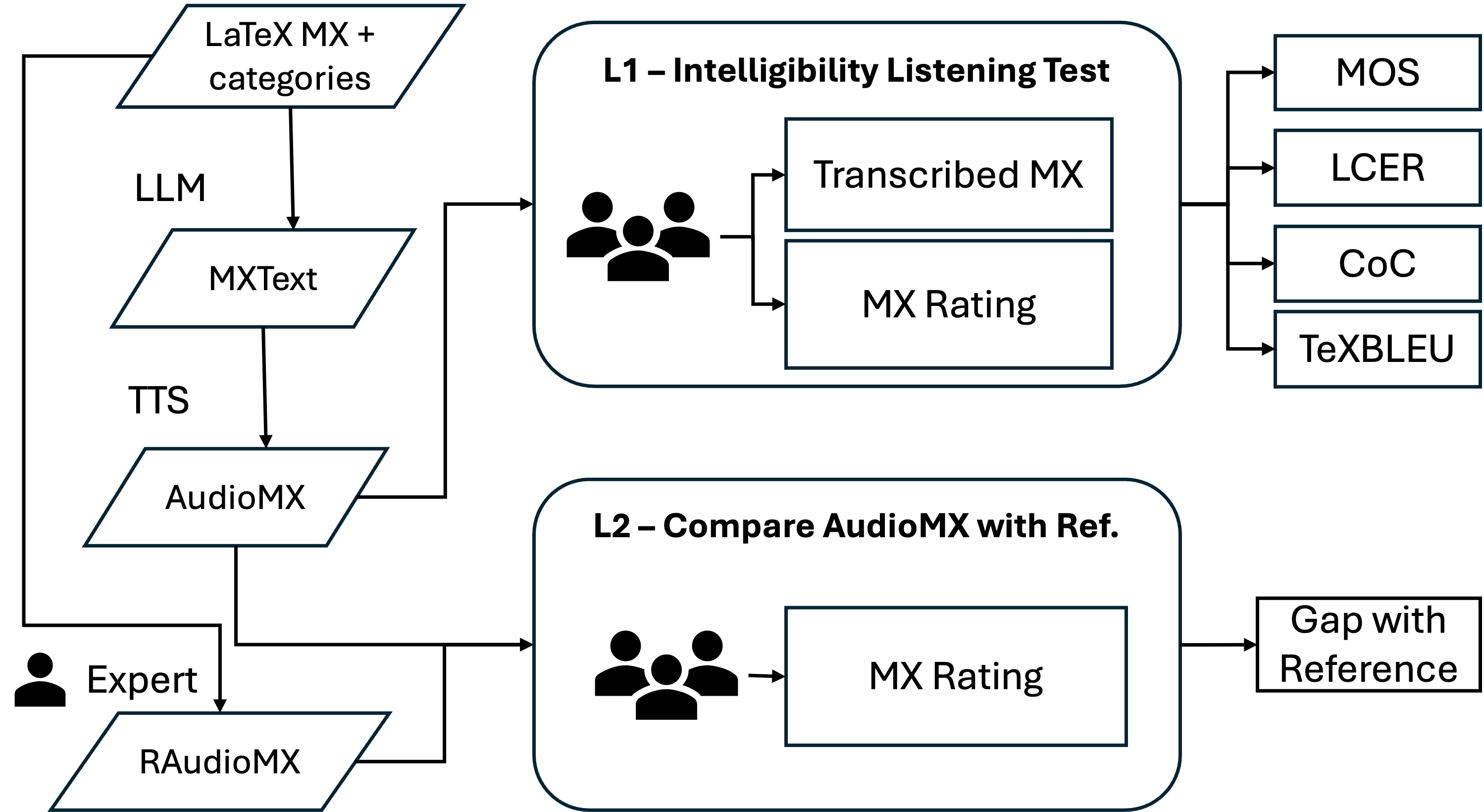}
    \caption{Dataset formats, experiment design and metrics.}
    \label{fig:exptdesign}
\end{figure}
\vspace{-0.2cm}
\section{Dataset and Methodology}
\label{sec:methodologyAndDataset}
\subsection{Dataset}\label{subsec:dataset}

There are limited datasets focusing on TTS using MX. HAMEX \cite{quiniou2011hamex} had handwritten MX and their audio recordings, was one of the earliest, but to the best of our knowledge, is no longer publicly available. The Kaggle dataset Handwritten Mathematical Equations (HME) \cite{handwritten_math_expressions}, a superset of HAMEX, has the handwritten MX and \LaTeX~ ground truth, but not audio files.

For this work, we consolidate MX from different folders of HME to include categories such as matrices, de-duplicate and obtain a dataset of 3141 MX. We manually categorize these into 8 different categories. Table \ref{tab:categories} lists the categories and a sample expression for each category. We sample a total of 120 expressions from the 8 categories for the listening tests. 
%MX containing matrices are not present in the HAMEX subset of HME, but are present in a similar format in a separate subfolder.  Post de-duplication, we consolidate 3,141 expressions across the HAMEX and Matrices subfolders of HME dataset.

Though \textit{Mathspeech} \cite{hyeon2024mathspeech} (comprising of audio clippings from YouTube videos of classrooms) has recently been released, it is primarily designed for ASR and was released after our research efforts. It also has fewer MX categories than HME.

\input{Dataset}
% Mathspeech \cite{hyeon2024mathspeech}, a very recent dataset of audio transcriptions for MX, has recently been released using audio clippings from YouTube videos of classrooms. This is primarily meant for Automatic Speech Recognition (ASR); in addition, Mathspeech research work and its associated dataset has been released after our research efforts. We also find that the Mathspeech data also has fewer varieties of equations than HME. 

% We use equations from the HAMEX subset and ``MatricesTrain" and ``MatricesTest" subfolders in HME dataset for our experiments. 

% \vspace{-1cm}
\vspace{-0.2cm}
\subsection{Methodology}\label{subsec:methodology}

We consider Seamless (SL) \cite{seamless2023}, an open-source TTS model, and four publicly available models via Application Program Interfaces (APIs) \textit{i.e.},  Google's Text-to-Speech AI API (GA) \cite{googleTTS}, Microsoft Azure AI Speech (AZ) \cite{azureAISpeech}, OpenAI Text-to-speech (OA) \cite{openaiTTS}, Amazon Polly (AP) \cite{amazonPolly}. The models we have chosen are among the SOTA TTS models \cite{LabelboxTTS2024}.
\vspace{-0.2cm}
\subsubsection{Direct processing of MX in TTS}\label{subsubsec:latex}

We provide the original \LaTeX~ expressions to the TTS models - however, none of the models perform well. After a manual evaluation for a few expressions across all models, we did not proceed with this approach.
\vspace{-0.2cm}
\subsubsection{Use of LLM for pronunciation}\label{subsubsec:llm}

To overcome the above challenge (Section \ref{subsubsec:latex}), we present the expressions to a LLM with the following prompt - ``\textit{Provide a pronunciation for the mathematical expression given in Latex below}''.  We did not explore chain of thoughts (CoT) and other advanced prompting techniques as we focus on evaluation of output of TTS systems for intelligibility. We focus on using an open-source model (QWEN2.5-7b) which is a relatively small LLM. However, in addition, for some of the expressions, we also provide them to GPT4 for comparison. Table \ref{tab:categories} shows the count of MX for which we got pronunciation from QWEN per category. The rest are from GPT4. This is an alternative to fine-tuning the T5 model step or LSTM models block in \cite{mathbridge,awais2024mathvision} respectively. Use of an existing LLM also reduces training costs. We call the generated pronunciation as MXText.% as against earlier works. }
\vspace{-0.2cm}
\subsubsection{Text to Speech Models}\label{subsubsec:tts}

 MXText is input to each TTS model in order to obtain the corresponding audio output (referred to as AudioMX). AudioMX are inputs to our listening tests to measure intelligibility and quality. %{\color{red} can we give the default parameters for each of these TTS models here? }
\vspace{-0.4cm}
\section{Experimental design} 
\label{sec:exp_design}
% For this research, we conduct two surveys with listeners
% \begin{itemize}
%     \item L1 test - intelligibility of the audio file without any additional information
%     \item L2 test - comparative ratings of the audio from listeners across different TTS outpuss with the MX displayed as a reference
% \end{itemize}

\input{combined_tables}
\color{black} Figure \ref{fig:exptdesign} shows our experimental design of two listening tests.
% described below. 
\vspace{-0.2cm}
\subsection{L1 - Intelligibility Listening Tests}\label{subsec:intelligibility}

{\color{black}To address RQ1, }the L1 listening test measures intelligibility of MX using only AudioMX. The listeners are not provided with the ground truth MX, the LLM or the TTS model corresponding to the AudioMX. The listeners are asked to transcribe the MX in \LaTeX~ to the best of their understanding from the AudioMX. The MX written in \LaTeX~ is rendered as preview immediately so that any correction can be made.
% ({\color{red}{screenshot attached in supplementary material}}). 
Listeners can play the audio as many times as required and/or change speed (0.5x to 2x of the actual speed). If unable to comprehend and/or transcribe the AudioMX, they were asked to enter ``\textit{Did not understand}''. The listeners also provide their opinion score on the audio based on intelligibility on a Likert scale of $1-5$ ($1$ being the poor and $5$ being the excellent). The audio rating is considered to be $1$ in case they could not understand the AudioMX. The Mean Opinion Score (MOS) from these ratings provides a subjective view of the perceptual difficulty in audio understanding. The listener transcribed \LaTeX\ MX provides an objective view of the intelligibility of the AudioMX. 

For L1 test, we sample 120 MX across 8 categories (Table \ref{tab:categories}). Each listener gets 2 questions   from the matrices category and 4 questions  from each of the other categories. With 5 TTS models per MX, we evaluate a total of 600 AudioMX. Each AudioMX is rated by 3 listeners - thus we obtain a total of 1800 opinion scores and listener transcribed \LaTeX~ MX.  %For all the categoriescategories other than Matrices, 3 of the answers were generated by QWEN and 1 by GPT4. ##commented by Ranjani .
The L1 test was conducted in 60 batches, each having 30 AudioMX. We enrolled 49 listeners for this, a few of who did at most 3 batches.  %who completed the survey for 60 users (2 individuals were requested to do for 2 users and 3 individuals were requested to do for 3 users). 
All of the listeners who participated in the listening tests have age range between 18-50 years, a background in science or engineering education and are familiar with \LaTeX. 
\vspace{-0.2cm}
\subsection{Evaluation of transcribed MX}\label{subsec:evaluation}

Based on challenges in evaluating MX \cite{jung2024texbleu, mathbridge}, we consider the following three approaches to compare transcribed MX with the ground truth MX. The results are shown in Table \ref{tab:L1scores}.
\begin{itemize}
    \item Four evaluators manually compare the ground truth MX and transcribed MX  marking them as correct or incorrect. The \LaTeX\ was rendered for ease of correction. Each MX was marked correct or wrong based on the equivalence of the relations. The evaluators do not have of any details of the response (such as the MX category, LLM, TTS model, listener identity who had transcribed the MX, or the original audio rating). 
% The evaluators were asked to make basic corrections to the \LaTeX\ for consistency so that the accuracy of the character-based error was not unfairly affected. For example, if a user had written \texttt{$3/5$} instead of \texttt{$\frac{3}{5}$}, the evaluator manually corrected it to the latter form to avoid an undue penalty in error calculation. Mathematical expressions were marked correct or wrong based on their mathematical equivalence. However, for the mathematical expression $x - \frac{y}{1} = 5$, if the user had written $\frac{x - y}{1} = 5$, the evaluator marked it as wrong since the mathematical equivalence holds only for the special case where the denominator is $1$. 
We call the metric derived out of this expert validation as Count-of-Correct (CoC). We do not increment the counter in case the response is ``\textit{Did not understand}''.
\item For evaluating how close the transcribed MX is to the original MX we use LCER. This is calculated by removing spaces from the \LaTeX\ representation and then computing the Levenshtein distance \cite{kukich1992techniques}. This is then normalized by min-max normalization to transform it in the range of $[0,1]$. In case the response is ``\textit{Did not understand}'', we set LCER to 1.
\item We compute TeXBLEU \cite{jung2024texbleu} to evaluate similarity between the ground truth and user-written MX. TeXBLEU captures structural and syntactic differences, providing a score between 0 and 1, higher values indicate greater similarity. For response ``\textit{Did not understand}'', the score was set to 0. %{\color{red}Where is this score in Table 2?}
\end{itemize}

% \input{cer_algo}
% This method complemented manual evaluation and character-level error analysis, ensuring a more comprehensive assessment of mathematical expression accuracy.

% \begin{figure}[h]
%     \centering
%     \includegraphics[width=\linewidth]{cer_and_correct_percentage.png}
%     \caption{CER and \% of correct across categories and models.}
%     \label{fig:enter-label}
% \end{figure}

% {\color{red} Nistha this required your algorithm. We also need a paragraph for TexBleu comparison}
% \input{UserScores}
\input{mushraScores}

\vspace{-0.7cm}
\subsection{L2 - Comparison of TTS AudioMX with Reference}\label{subsec:mushra}

{\color{black}To address RQ2, }the L2 listening test is designed to measure user preference of different TTS models along with a hidden reference AudioMX. This provides an indication of the gap with SOTA models for MX. 4 experts are asked to record the MX in a way they would typically communicate to an audience via audio alone. This is considered as Reference AudioMX (RAudioMX).

The L2 comparative tests are motivated by MUlti Stimulus test with Hidden Reference and Anchor (MUSHRA) recommendations \cite{ITU-R-BS1534}. Instead of explicit reference audio, we render the MX visually \footnote{Screenshot of the listening test UI in supplementary material}.  
Thus for every MX, listeners are asked to rate the TTS AudioMX and (hidden) RAudioMX on a scale of [0,100] with respect to clarity and intelligibility of the displayed MX.  The details of the TTS model or LLM corresponding to the AudioMX are not provided. 
% Although it was not mentioned in the test, the listeners could identify the reference because the human rated audio sounded different and had a separate accent with respect to the TTS model generated audios.
Each MX is rated by 3 listeners. We sample $35$ MX from $5$ categories (excluding matrices).  
We split the MX into 15 batches with each listener rating 7 MX (one from each category). Each MX comparison has 6 audio files, including the hidden reference, resulting in 42 score preferences. L2 test is completed by 13 individuals with 2 individuals scoring at most 2 batches each.
\vspace{-0.2cm}
\section{Results and Analysis} 
\label{sec:results}

Table \ref{tab:L1scores} tabulates the average MOS, LCER, CoC and TexBLEU metrics (detailed in Section \ref{subsec:intelligibility} and \ref{subsec:evaluation}) across categories and the considered TTS models for L1 listening test. These metrics address \textbf{RQ1}. From Table \ref{tab:L1scores}(a)-(d), we observe that there is no TTS model which scores consistently high across categories and across metrics. It is also interesting to note that models having better CoC, lower LCER and high TeXBLEU score imply consistent better MOS scores and vice-versa. Hence to identify the significant main and interactions effects among the factors `LLM', `TTS Model' and `Category' from the L1 scores, we perform An Analysis of Variance (ANoVA) \cite{st1989analysis} tests on LCER scores where `TTS Model' and `Category' factors are found to be statistically significant ($p$ value $< 0.05$). We infer that comprehension of MX from a TTS audio depends on category of MX apart from the TTS model itself. Although ANoVA with MOS as the response additionally identifies factor `LLM' also as significant, we conclude that the listener rating of AudioMX is not reflected in their MX transcription. Further, we manually check the correctness of pronunciations (87.5\% overall, 85.9\% for QWEN, 92.9\% for GPT4) and perform ANoVA with correctly pronounced MX. We observe that `LLM' becomes an insignificant factor with MOS as the response, but there is no change of inference in the study with LCER as the response. Thus, when the pronunciation is correct, LCER and MOS are statistically equivalent across both LLMs, irrespective of the LLM size.

We also analyse CoC considering length of the audio; we observe that audio which are relatively long are more difficult to transcribe. {\color{black} The CoC was at 64\% for AudioMX having duration $<$ 3s, this reduces to 57.1\% considering AudioMX of upto 6s duration and 55.0\% those upto 12s duration. 97.5\% of AudioMX are less than 12s duration, with the longest being 18.2s. }

\input{pronunciation}

%% A careful statistical analysis shows that the relative ordering via MOS and CER and CoC are not necessarily fully in sync. For example for although statistical analysis of MOS shows that all the TTS models are equal,  CER shows $AP=AZ > SL=GA=OA$ and CoC shows $AP=AZ=OA > GA=SL$ where the equality sign indicates that the models are equal statistically and greater sign indicates that one model being better than the other. On the other hand for complex equations like those in the ``Summation'' category, CER indicates $AZ > AP=GA=OA > SL$ although MOS shows $OA=GA=AZ=AP > SL$. This indicates there is a gap between user's perception of them being able to follow an MX and their ability to correctly transcribe it.
% \vspace{-0.7cm}
For RQ2, we use the human reference audios - they have average CoC is 97\% across categories, average LCER is 0.01 with only one of 35 equations being incorrectly transcribed. From L2 listening test scores (detailed in Section \ref{subsec:mushra}), we calculate the difference between the listener score for RAudioMX and every TTS AudioMX for each MX. The mean of these per category are presented in Table \ref{tab:mushraScores}. For all MX, the score for (hidden) RAudioMX is highest. We observe the scores to vary across categories in the results for this test too.  Our results show that for MX category `Numerics', all the models are scored close to reference; for `Fractions', all models except SL are close to reference. In all other categories, there exists large gaps with respect to the reference audio. The gap observed in Table \ref{tab:mushraScores} implies that public API or open source general purpose TTS models do not include the complexities of prosody particular to MX and is in line with discussions in \cite{souza2020towards}. 

 In conclusion, for \textbf{RQ1}, we observe that outputs for MX of SOTA TTS models  are not necessarily intelligible to a human listener. The gap in intelligibility varies across categories and models. For \textbf{RQ2}, we establish that for most categories there is a large gap with respect to the reference indicating a significant scope for improvement in TTS for MX even for SOTA models.

\section{Discussion} 
\label{sec:discussion}
% \subsection{Evaluation Challenges - {\color{red} CAN BE DROPPED}}\label{subsec:texBleu}

% One major challenge for evaluating \LaTeX\ is that similar expressions have equivalent but different representations, e.g., via \textit{pmatrix} vs. \textit{bmatrix}, using \texttt{4/5} instead of \(\frac{4}{5}\) - this is why we overcome this by looking at a host of metrics like CoC, LCER and TeXBLEU. 
% % {\color{red} need 2-3 examples of TexBleu fail, as well as CER fail}

\subsection{Qualitative Analysis of MX Transcription Error}\label{subsec:qualAnalysis}

Some of the patterns and errors observed during process of manual verification of transcribed MX (discussed in Section \ref{subsec:evaluation}), common feedbacks resulting from a focus-group discussion with listeners are detailed below.
\begin{itemize}
\item Pronunciation related 
\begin{itemize}
    \item {\color{black} {Confusion among variables such as (`b', `p'), similar words such as ($\pi$, five, $\phi$) has resulted in higher LCER, reduced CoC and TeXBLEU metrics.}
    \item Variations of TTS outputs of `cos t' as \textipa{[koh.saIn ti]}, \textipa{[kohs ti]}; incorrect pronunciation TTS outputs of `sin t' as \textipa{[sIn ti]} instead of \textipa{[saIn ti]} has resulted in reduced MOS Scores.} 
\end{itemize}
\item Prosody related
\begin{itemize}
    \item For many MX there was a lack of clarity on the parenthesis or order of operations implied in the MX. For example, the MX $\frac{a^2-b^2}{a}$ might be transcribed as $a^2-\frac{b^2}{a}$; similarly $50 \pm \frac{51}{144}$ with some transcribing it as $\frac{50\pm51}{144}$. These observations are in line with those in \cite{korzh2024listening}.
    \item Varied pauses (across TTS models) between words `square' and `root' can result in MX $(-a-b\sqrt{2})$ to be comprehended and transcribed as $(-a-b^2\sqrt{2})$. These increase LCER and reduce CoC and TeXBLEU metrics. 
    \item The TTS models do not consider appropriate pauses nor adjust the speed of audio for more complex MX (such as calculus, summation, or those having equality signs). This results in multiple listenings per MX, causing listening fatigue. The incorrect speed and pause results in perception that AudioMX is unclear and sometimes, rushed; for example, there was no pause between `integral from zero to infinity' and the variables within integral; these are typical suprasegmental prosodic components of RAudioMX.
    \item Changing the AudioMX speed was possible but it did not necessarily result in better comprehension of MX. 
    %\item {\color{blue} Some models do not necessarily synthesize the pronounciation provided via MX Text. For example, AZ AudioMX for $\cos$ is `cosine' although the pronounciation text (MX text) has only cos. In the L2 experimental setup, it also has better prosody for calculus based equations leading to significantly lower gap with reference in Table \ref{tab:mushraScores}}.
    %\item {\color{blue} The gap observed in Table \ref{tab:mushraScores} implies that public API or open source general purpose TTS models do not include the complexities of prosody particular to MX as discussed in \cite{souza2020towards}}
    \item Listeners found some of the AudioMX to be longer in duration and hence more challenges in transcribing. This is in line with the duration analysis in Section \ref{sec:results}.
\end{itemize}
    
    \item Comprehending AudioMX involving matrices and summation was particularly difficult and many listeners marked them as `Did not understand'.  
\end{itemize}

% are (i) Listeners felt that AudioMX was often rushed and there were no significant pauses between operands  This indicated that the TTS models were not able to provide the correct prosody for the Audio MX and had little or no idea of the expression that the model was transcribing. 

\subsection{Differences with reference}\label{subsec:reference}

As mentioned in Section \ref{sec:exp_design}, RAudioMX is recorded  without using LLM-generated pronunciation. Most experts used descriptors like `whole divided by' for parenthesis e.g. `a-squared minus b-squared \textbf{whole divided by} a' for $\frac{a^2-b^2}{a}$. In addition, appropriate pauses between operands, variables, equality signs are introduced. The scores from L2 listening test indicate that these greatly affect the scoring of an AudioMX to the listener. 
Prosody for MX therefore needs comprehension of the expression; this is absent across all TTS models except for the simplest of categories such as numerics and fractions.

\subsection{Automatic Speech Recognition Check}\label{subsec:asrCheck}

{\color{black} The AudioMX from TTS is passed through ASR model (we choose Whisper-base model \cite{radford2023robust}). We then compute ASR metrics from this cascaded system - Character Error Rate (CER), Bilingual Evaluation Understudy (BLEU) and Recall-Oriented Understudy for Gisting Evaluation (ROUGE) scores \cite{labied2024assessing} w.r.t. MXText. Average CER is 0.10, BLEU is 0.66 and ROUGE is 0.84 indicating an faithful pipeline, averaged across all TTS models. The correlation of the ASR based metrics (CER, BLEU and ROUGE) with the transcription based metrics (CoC, LCER, TeXBLEU) is very low -  the absolute correlation value between any ASR metric and any transcription based metrics is less than 0.15. Thus, whilst the TTS-ASR cascaded models can result in high overall ASR metrics w.r.t. MXText, they are not necessarily intelligible to a human listener.}

\section{Conclusions \& Future Work}

% In this study, we considered two RQs around the intelligibility of AudioMX using publicly available SOTA APIs and one open source model. For RQ1, we establish that for most categories, human listeners experience significant challenges in being able to transcribe from AudioMX and have poor MOS. For RQ2, we demonstrate that there are gaps wrt RAudioMX for all TTS models and most MX categories. 

Our research establishes the variability of TTS performance (for \textbf{RQ1}) based on category and the gap with human rendered audio (for \textbf{RQ2}). We observe that LLMs, such as GPT4, can provide the pronunciations from MX. Any real-world system will need an analysis of the categories of expressions relevant in that application; the choice of TTS models will be based on relative performance for these categories. One may need custom-trained / fine-tuned TTS models because of lack of prosody for MX in the existing publicly available TTS models.

Future work involves fine-tuning models to improve performance on audio rendering of MX and evaluating them with current baselines.

\bibliographystyle{IEEEtran}
\bibliography{mybib}

\end{document}

%% file: Dataset.tex
\begin{table}[h]
  \centering
  \resizebox{0.49\textwidth}{!}{
  \begin{tabular}[width=\textwidth]{lcccc}
    \toprule
    Category & No. of expressions & QWEN  & Sample expression \\
    \midrule
    Advanced Algebraic & 16 & 12  & 
    $10x^2\sqrt{5+2\sqrt{5}} + 5x - \sqrt{5+2\sqrt{5}}$ \\
    Basic Algebraic & 16 & 12  & 
    $z-2$ \\
    Fractions & 16 & 12  & 
    $\frac{1}{r^n}$ \\
    Calculus & 16 & 12  & 
    $\int \cos t \, dt = \sin t$ \\
    Matrices & 8 & 8  & 
    $\begin {pmatrix} 1 & 0 \\ 0 & 1 \end {pmatrix}$ \\
    Numerics & 16 & 12  & 
    $14 + 61 = 75$ \\
    Sq. Roots and Trig & 16 & 12  & 
    $e^{\alpha t}\cos\beta t$ \\
    Summation & 16 & 12  & 
    $\sum_{i=1}^{36} i = 666$ \\
    \midrule
    Total & 120 & 92  & \\
    \bottomrule
  \end{tabular}
  }
  \caption{Categories used in our experiments along with a sample expression. The column QWEN denotes number of pronunciations from QWEN model, rest are from GPT4}
  \label{tab:categories}
\end{table}

%% file: combined_tables.tex
% Please add the following required packages to your document preamble:
% \usepackage{multirow}
\begin{table*}[t]
\centering
\scalebox{0.75}{

\begin{tabular}{lrrrrrlrrrrrlrrrrrlrrrrr}
\cline{1-6} \cline{8-12} \cline{14-18} \cline{20-24}
\multicolumn{1}{|c|}{\textbf{Metric}}    & \multicolumn{5}{c|}{\textbf{Mean Opinion Scores (MOS)}}                                                                              & \multicolumn{1}{l|}{} & \multicolumn{5}{c|}{\textbf{Count of Correct (\%)}}                                                                                       & \multicolumn{1}{l|}{} & \multicolumn{5}{c|}{\textbf{\LaTeX~ Character Error Rate}}                                                                             & \multicolumn{1}{l|}{} & \multicolumn{5}{c|}{\textbf{TeXBLEU}}                                                                                                \\ \cline{1-6} \cline{8-12} \cline{14-18} \cline{20-24} 
\multicolumn{1}{|c|}{Category}           & \multicolumn{1}{c|}{AP}  & \multicolumn{1}{c|}{AZ}  & \multicolumn{1}{c|}{GA}  & \multicolumn{1}{c|}{OA}  & \multicolumn{1}{c|}{SL}  & \multicolumn{1}{l|}{} & \multicolumn{1}{c|}{AP}   & \multicolumn{1}{c|}{AZ}   & \multicolumn{1}{c|}{GA}   & \multicolumn{1}{c|}{OA}   & \multicolumn{1}{c|}{SL}   & \multicolumn{1}{l|}{} & \multicolumn{1}{c|}{AP}  & \multicolumn{1}{c|}{AZ}  & \multicolumn{1}{c|}{GA}  & \multicolumn{1}{c|}{OA}  & \multicolumn{1}{c|}{SL}  & \multicolumn{1}{l|}{} & \multicolumn{1}{c|}{AP}  & \multicolumn{1}{c|}{AZ}  & \multicolumn{1}{c|}{GA}  & \multicolumn{1}{c|}{OA}  & \multicolumn{1}{c|}{SL}  \\ \cline{1-6} \cline{8-12} \cline{14-18} \cline{20-24} 
\multicolumn{1}{|l|}{Advanced Algebraic} & \multicolumn{1}{r|}{3.9} & \multicolumn{1}{r|}{3.5} & \multicolumn{1}{r|}{\textbf{4.3}} & \multicolumn{1}{r|}{3.9} & \multicolumn{1}{r|}{3.1} & \multicolumn{1}{l|}{} & \multicolumn{1}{r|}{33.3} & \multicolumn{1}{r|}{39.6} & \multicolumn{1}{r|}{\textbf{52.1}} & \multicolumn{1}{r|}{\textbf{52.1}} & \multicolumn{1}{r|}{33.3} & \multicolumn{1}{l|}{} & \multicolumn{1}{r|}{0.1} & \multicolumn{1}{r|}{0.1} & \multicolumn{1}{r|}{\textbf{0.0}} & \multicolumn{1}{r|}{0.1} & \multicolumn{1}{r|}{0.2} & \multicolumn{1}{l|}{} & \multicolumn{1}{r|}{\textbf{0.9}} & \multicolumn{1}{r|}{0.8} & \multicolumn{1}{r|}{\textbf{0.9}} & \multicolumn{1}{r|}{\textbf{0.9}} & \multicolumn{1}{r|}{0.8} \\ %\cline{1-6} \cline{8-12} \cline{14-18} \cline{20-24} 
\multicolumn{1}{|l|}{Basic Algebraic}    & \multicolumn{1}{r|}{\textbf{4.0}} & \multicolumn{1}{r|}{3.9} & \multicolumn{1}{r|}{3.6} & \multicolumn{1}{r|}{3.5} & \multicolumn{1}{r|}{3.2} & \multicolumn{1}{l|}{} & \multicolumn{1}{r|}{\textbf{70.8}} & \multicolumn{1}{r|}{\textbf{70.8}} & \multicolumn{1}{r|}{47.9} & \multicolumn{1}{r|}{66.7} & \multicolumn{1}{r|}{33.3} & \multicolumn{1}{l|}{} & \multicolumn{1}{r|}{\textbf{0.1}} & \multicolumn{1}{r|}{\textbf{0.1}} & \multicolumn{1}{r|}{0.2} & \multicolumn{1}{r|}{0.2} & \multicolumn{1}{r|}{\textbf{0.1}} & \multicolumn{1}{l|}{} & \multicolumn{1}{r|}{\textbf{0.9}} & \multicolumn{1}{r|}{\textbf{0.9}} & \multicolumn{1}{r|}{0.8} & \multicolumn{1}{r|}{0.8} & \multicolumn{1}{r|}{0.8} \\ %\cline{1-6} \cline{8-12} \cline{14-18} \cline{20-24} 
\multicolumn{1}{|l|}{Calculus}           & \multicolumn{1}{r|}{3.6} & \multicolumn{1}{r|}{3.6} & \multicolumn{1}{r|}{3.7} & \multicolumn{1}{r|}{\textbf{3.9}} & \multicolumn{1}{r|}{2.6} & \multicolumn{1}{l|}{} & \multicolumn{1}{r|}{54.2} & \multicolumn{1}{r|}{50.0} & \multicolumn{1}{r|}{39.6} & \multicolumn{1}{r|}{\textbf{60.4}} & \multicolumn{1}{r|}{18.8}  & \multicolumn{1}{l|}{} & \multicolumn{1}{r|}{\textbf{0.1}} & \multicolumn{1}{r|}{\textbf{0.1}} & \multicolumn{1}{r|}{\textbf{0.1}} & \multicolumn{1}{r|}{\textbf{0.1}} & \multicolumn{1}{r|}{0.3} & \multicolumn{1}{l|}{} & \multicolumn{1}{r|}{\textbf{0.8}} & \multicolumn{1}{r|}{\textbf{0.8}} & \multicolumn{1}{r|}{0.7} & \multicolumn{1}{r|}{\textbf{0.8}} & \multicolumn{1}{r|}{0.6} \\ %\cline{1-6} \cline{8-12} \cline{14-18} \cline{20-24} 
\multicolumn{1}{|l|}{Fractions}          & \multicolumn{1}{r|}{4.1} & \multicolumn{1}{r|}{\textbf{4.3}} & \multicolumn{1}{r|}{4.2} & \multicolumn{1}{r|}{4.1} & \multicolumn{1}{r|}{3.9} & \multicolumn{1}{l|}{} & \multicolumn{1}{r|}{58.3} & \multicolumn{1}{r|}{58.3} & \multicolumn{1}{r|}{\textbf{66.7}} & \multicolumn{1}{r|}{62.5} & \multicolumn{1}{r|}{47.9} & \multicolumn{1}{l|}{} & \multicolumn{1}{r|}{0.1} & \multicolumn{1}{r|}{\textbf{0.0}} & \multicolumn{1}{r|}{0.1} & \multicolumn{1}{r|}{\textbf{0.0}} & \multicolumn{1}{r|}{0.1} & \multicolumn{1}{l|}{} & \multicolumn{1}{r|}{\textbf{0.8}} & \multicolumn{1}{r|}{\textbf{0.8}} & \multicolumn{1}{r|}{\textbf{0.8}} & \multicolumn{1}{r|}{\textbf{0.8}} & \multicolumn{1}{r|}{\textbf{0.8}} \\ %\cline{1-6} \cline{8-12} \cline{14-18} \cline{20-24} 
\multicolumn{1}{|l|}{Matrices}           & \multicolumn{1}{r|}{3.1} & \multicolumn{1}{r|}{3.1} & \multicolumn{1}{r|}{3.1} & \multicolumn{1}{r|}{3.5} & \multicolumn{1}{r|}{\textbf{3.6}} & \multicolumn{1}{l|}{} & \multicolumn{1}{r|}{20.8} & \multicolumn{1}{r|}{16.7} & \multicolumn{1}{r|}{25.0} & \multicolumn{1}{r|}{\textbf{29.2}} & \multicolumn{1}{r|}{20.8} & \multicolumn{1}{l|}{} & \multicolumn{1}{r|}{0.2} & \multicolumn{1}{r|}{0.2} & \multicolumn{1}{r|}{0.2} & \multicolumn{1}{r|}{\textbf{0.1}} & \multicolumn{1}{r|}{\textbf{0.1}} & \multicolumn{1}{l|}{} & \multicolumn{1}{r|}{0.7} & \multicolumn{1}{r|}{0.7} & \multicolumn{1}{r|}{0.7} & \multicolumn{1}{r|}{\textbf{0.8}} & \multicolumn{1}{r|}{\textbf{0.8}} \\ %\cline{1-6} \cline{8-12} \cline{14-18} \cline{20-24} 
\multicolumn{1}{|l|}{Numerics}           & \multicolumn{1}{r|}{\textbf{4.9}} & \multicolumn{1}{r|}{4.8} & \multicolumn{1}{r|}{\textbf{4.9}} & \multicolumn{1}{r|}{\textbf{4.9}} & \multicolumn{1}{r|}{4.7} & \multicolumn{1}{l|}{} & \multicolumn{1}{r|}{97.9} & \multicolumn{1}{r|}{97.9} & \multicolumn{1}{r|}{97.9} & \multicolumn{1}{r|}{\textbf{100.0}} & \multicolumn{1}{r|}{\textbf{100.0}} & \multicolumn{1}{l|}{} & \multicolumn{1}{r|}{\textbf{0.0}} & \multicolumn{1}{r|}{\textbf{0.0}} & \multicolumn{1}{r|}{\textbf{0.0}} & \multicolumn{1}{r|}{\textbf{0.0}} & \multicolumn{1}{r|}{\textbf{0.0}} & \multicolumn{1}{l|}{} & \multicolumn{1}{r|}{\textbf{1.0}} & \multicolumn{1}{r|}{\textbf{1.0}} & \multicolumn{1}{r|}{\textbf{1.0}} & \multicolumn{1}{r|}{\textbf{1.0}} & \multicolumn{1}{r|}{\textbf{1.0}} \\ %\cline{1-6} \cline{8-12} \cline{14-18} \cline{20-24} 
\multicolumn{1}{|l|}{Sq. Roots and Trig} & \multicolumn{1}{r|}{3.6} & \multicolumn{1}{r|}{4.0} & \multicolumn{1}{r|}{\textbf{4.4}} & \multicolumn{1}{r|}{4.1} & \multicolumn{1}{r|}{3.3} & \multicolumn{1}{l|}{} & \multicolumn{1}{r|}{50.0} & \multicolumn{1}{r|}{54.2} & \multicolumn{1}{r|}{\textbf{64.6}} & \multicolumn{1}{r|}{58.3} & \multicolumn{1}{r|}{29.2} & \multicolumn{1}{l|}{} & \multicolumn{1}{r|}{0.1} & \multicolumn{1}{r|}{0.1} & \multicolumn{1}{r|}{\textbf{0.0}} & \multicolumn{1}{r|}{0.1} & \multicolumn{1}{r|}{0.1} & \multicolumn{1}{l|}{} & \multicolumn{1}{r|}{\textbf{0.9}} & \multicolumn{1}{r|}{0.8} & \multicolumn{1}{r|}{\textbf{0.9}} & \multicolumn{1}{r|}{\textbf{0.9}} & \multicolumn{1}{r|}{0.8} \\ %\cline{1-6} \cline{8-12} \cline{14-18} \cline{20-24} 
\multicolumn{1}{|l|}{Summation}          & \multicolumn{1}{r|}{3.0} & \multicolumn{1}{r|}{3.3} & \multicolumn{1}{r|}{\textbf{3.5}} & \multicolumn{1}{r|}{\textbf{3.5}} & \multicolumn{1}{r|}{2.3} & \multicolumn{1}{l|}{} & \multicolumn{1}{r|}{41.7} & \multicolumn{1}{r|}{\textbf{58.3}} & \multicolumn{1}{r|}{45.8} & \multicolumn{1}{r|}{45.8} & \multicolumn{1}{r|}{39.6}  & \multicolumn{1}{l|}{} & \multicolumn{1}{r|}{0.2} & \multicolumn{1}{r|}{\textbf{0.1}} & \multicolumn{1}{r|}{0.2} & \multicolumn{1}{r|}{0.2} & \multicolumn{1}{r|}{0.4} & \multicolumn{1}{l|}{} & \multicolumn{1}{r|}{0.7} & \multicolumn{1}{r|}{\textbf{0.8}} & \multicolumn{1}{r|}{0.7} & \multicolumn{1}{r|}{0.7} & \multicolumn{1}{r|}{0.5} \\ \cline{1-6} \cline{8-12} \cline{14-18} \cline{20-24} 
                                         & \multicolumn{1}{l}{}     & \multicolumn{1}{l}{}     & \multicolumn{1}{l}{}     & \multicolumn{1}{l}{}     & \multicolumn{1}{l}{}     &                       & \multicolumn{1}{l}{}      & \multicolumn{1}{l}{}      & \multicolumn{1}{l}{}      & \multicolumn{1}{l}{}      & \multicolumn{1}{l}{}      &                       & \multicolumn{1}{l}{}     & \multicolumn{1}{l}{}     & \multicolumn{1}{l}{}     & \multicolumn{1}{l}{}     & \multicolumn{1}{l}{}     &                       & \multicolumn{1}{l}{}     & \multicolumn{1}{l}{}     & \multicolumn{1}{l}{}     & \multicolumn{1}{l}{}     & \multicolumn{1}{l}{}     \\
\multicolumn{1}{c}{}                     & \multicolumn{5}{c}{(a) MOS}                                                                                                          & \multicolumn{1}{c}{}  & \multicolumn{5}{c}{(b) CoC as \%}                                                                                                         & \multicolumn{1}{c}{}  & \multicolumn{5}{c}{(c) LCER}                                                                                                         & \multicolumn{1}{c}{}  & \multicolumn{5}{c}{(d) TeXBLEU}                                                                                                     
\end{tabular}

}
\caption{Average MOS, LCER, CoC (in \%) and TeXBLEU metrics across categories and TTS models for L1-listening test. Highlighted numbers indicate best average scores across models. \label{tab:L1scores}}
\end{table*}

%% file: mushraScores.tex
\begin{table}[h!]
\centering
\resizebox{0.4\textwidth}{!}{ 
\begin{tabular}{|l|r|r|r|r|r|}
\hline
Category                      & AP    & AZ    & GA    & OA    & SL     \\ \hline
Advanced Algebraic            & 28.1 & 34.1 & \textbf{24.4} & 24.7 & 51.4 \\ 
Basic Algebraic               & 26.7 & \textbf{24.1} & 36.9 & 25.6 & 47.3 \\ 
Calculus   & 30.9 & \textbf{9.5}  & 24.1 & 17.7 & 66.9 \\ 
Fractions                     & 4.6  & \textbf{2.7}  & 4.7  & 6.9  & 23.2 \\ 
Numerics                      & 3.0  & \textbf{1.45}  & 2.8  & 2.7  & 6.2  \\ 
Sq. Roots and Trig & 33.5 & 49.5 & \textbf{17.1} & 21.9 & 56.9 \\ 
Summation                     & 42.1 & 39.9 & \textbf{35.5} & 36.9 & 72.5 \\ \hline
\end{tabular}
}
\caption{Average of difference of reference score with TTS score for different categories in L2 listening test (lower the better).}
\label{tab:mushraScores}
\end{table}

%% file: pronunciation.tex
\begin{table}[t!]
\centering
\scalebox{0.85}{
\begin{tabular}{|p{0.2\linewidth}|p{0.4\linewidth}|p{0.38\linewidth}|} % Ensures text wrapping
    \hline
    Equation & LLM Pronunciation & Expected \\
    \hline
    $ \int\limits_{0}^{\infty} \frac{\sin^{2} x}{x^{2}} \,dx = \frac{\pi}{2}$ 
    & Integral from 0 to infinity of (sin x {\color{red}squared} over x squared) equals pi over 2 &  Integral from zero to infinity of, sin  squared x, over x squared, equals pi over 2 \\
    \hline
    $\begin{bmatrix} 1 & 0 \\ 0 & 1 \end{bmatrix}$
    & Pronounced as "one by one identity matrix" or simply "identity matrix." & Two by two identity matrix\\
    \hline
    $2^n - i$ & Two to the n minus i {\color{red}power} & Two to the n, minus i\\
    \hline
\end{tabular}
}
\caption{Representative pronunciation errors (marked in {\color{red}red}) identified by an expert and expected pronounciation.}
\label{tab:pronError}
\end{table}
\vspace{-0.3cm}